\documentclass[11pt,
showpacs,
preprintnumbers,amsmath,amssymb,
aps,prd,nofootinbib,eqsecnum,a4paper]{revtex4}
\usepackage{epsfig}
\usepackage[autostyle]{csquotes}
\usepackage{graphicx,epsf}
\usepackage{color}
\usepackage{bm}
\usepackage{psfrag}
\usepackage[symbol]{footmisc}
\usepackage{tensor}
\usepackage{caption}
\usepackage{subcaption}
\usepackage{hyperref}
\usepackage{amsmath}
\def\beq{\begin{equation}}
\def\eeq{\end{equation}}
\def\bea{\begin{eqnarray}}
\def\eea{\end{eqnarray}}
\def\bi{\begin{itemize}}
\def\ei{\end{itemize}}
\def\cs2{c_{\rm{s}}^2}
\def \beg {\begin{enumerate}}
\def \en {\end{enumerate}}

\def\M0{{\cal M}_0}

\begin{document}
%%%%%%%%%%%%%%%%%%%%%%

\title{An early and late times dynamical analysis of a scale invariant gravitational model with a vector scalar interaction: the isotropic case}

\author{R. Gonzalez Quaglia\footnote[1]{\href{mailto:rodrigo@icf.unam.mx}{rodrigo@icf.unam.mx}}}

\affiliation{
Instituto de Ciencias F\'{i}sicas, Universidad Nacional Aut\'{o}noma de M\'{e}xico,\\Av. Universidad S/N. Cuernavaca, Morelos, 62251, M\'{e}xico}
%
% \date{\today}

\begin{abstract}
Scalar fields are widely and popularly used in cosmology in order to explain different phenomena among which, inflation and dark energy are two of the most popular ones. Specifically, in recent years, scale invariance in the gravitational sector has gained interest due to its simplicity, ability to model inflation and the dynamical generation of the Planck scale. In this paper, motivated by a non minimally coupled scale invariant $R^{2}$ gravitational model originally proposed by  M.Rinaldi and L. Vanzo, we investigate how the inclusion of a vector scalar interaction that respects the scale invariance of the original model may affect the early and late time dynamics. We employ dynamical analysis tools in order to find the fixed points of the system and the local solutions for each variable around each fixed point finding out that the early universe solution of the extended model is compatible with that found in the original model with the exception of a new unstable fixed point appearing in the extended model. This new fixed point however has the same linealized solutions as the unstable fixed point found for the original model. We later employ numerical calculations in order to check that the analytical approach holds. Furthermore we show how the dynamical generation of the Planck mass is unaffected by the new field content of the model. Finally, we motivate and investigate a reduced version of the extended model finding out that, at late times, the extended model has only stable de Sitter fixed points where the scalar field becomes constant taking the role of the cosmological constant and the vector field is washed out.

\end{abstract}

% \pacs{98.80.Cq \hfill  arXiv:2008.07359}

\maketitle

%%%%%%%%%%%%%%%%%%%%%%%%%%%%%%%%
\section {\bf Introduction}\label{sec:intro}
\subsection{Motivation on scale invariance in gravity and inflation }
General Relativity (GR) \cite{The Foundation of the General Theory of Relativity} is the go-to theory when talking about gravity. It is the most complete and consistent gravitational theory we possess until today. From cosmological to galactic and even solar system scales GR has yet to be in disagreement with experiments. However, as every physical theory has a range of applicability, this theory is unable to describe, for example, the singularity at the center of a black hole, the initial singularity of the universe, and most important for us in this paper, GR together with the matter content of the $\Lambda$CDM model is incapable of describing and accelerated universe that can terminate\footnote{Inflation needs to terminate in order to give rise to the radiation dominated universe.} like the hypothesized inflationary epoch in the early universe\footnote{Note that pure GR with a cosmological constant indeed leads to an accelerated universe that will forever accelerate.}. In fact, because of this incapability of describing accelerated spacetimes, gravitational models which included scalar fields were developed, either to describe dark energy \cite{Dynamics of dark energy}-\cite{An Alternative to quintessence} or to model inflation \cite{Cosmological Consequences of a Rolling Homogeneous Scalar Field}-\cite{Guth}. The usual formulation of a model for inflation employs a minimally coupled scalar field which is the responsible of the acceleration of the universe and that naturally gives rise to a condition for this accelerated expansion to end.   
After this single field models were developed, observations started their golden age and therefore a way to rule out inflationary models was possible. This led to models of the type $V\propto \phi^{n}$ not being in agreement with observations \cite{Planck 2018 results. X. Constraints on inflation}. However, one model, the Starobinsky model \cite{Starobinsky} is, up until today, one of the models that best fit the observations from Planck satellite \cite{Planck 2018 results. X. Constraints on inflation}. The Starobinsky model is quite a different type of inflationary model because it works with a different type of gravitational action\footnote{Although it is true that the Starobinsky model can be recasted in an Einstein Hilbert plus scalar field form.}. This model is one kind of the so called $f(R)$ theories \cite{f(R)} which action reads 

\begin{equation}\label{Starobinsky}
    S=\frac{1}{2}\int d^{4}x\sqrt{-g}\left[\alpha R^{2}+R\right].
\end{equation}

Where $R$ is the Ricci scalar, $\alpha$ is a coupling constant and $g$ is the determinant of the metric. 
From this action we can also start the discussion and motivation on scale invariance in the gravitational sector. As already mentioned, the pure Einstein Hilbert term is incapable of giving birth to the accelerated expansion, hence the term $R^{2}$ must be the responsible of said accelerated expansion. Moreover, as it is thought that inflation takes place at very high energies, then, as the quadratic term in the Ricci scalar has units of $[E^{4}]$ it would seem that at very high energies, where the dominant term is the quadratic one, gravity is scale invariant. A second argument in favor of a scale invariant theory of gravitation is the fact that, as GR tends to be replaced with a modified theory of gravity in order to explain inflation, this modified theory must reproduce GR at the end of inflation for the $\Lambda$CDM model to hold. Typically, scale invariant models have the capability of giving birth to the Planck mass in a dynamical way \cite{Racioppi}-\cite{Scale invariance}
and so, these scale invariant theories are indeed capable of recovering GR at lower energies via spontaneous symmetry breaking around a stable minimum. Finally, this symmetry breaking feature could also be used as a way to couple the Standard Model of Particle Physics (SM) to gravity. As the SM is scale invariant whenever the Higgs mass is zero, it is natural to think that this mass may be born from the breaking of this scale invariant symmetry.
The organization of the paper is as follows: In section \ref{Addition of a vector field} we motivate the addition of vector fields to the model in \cite{Rinaldi}, we show the basic equations for this new vector field and we focus on the incompatibility between the vector field and the isotropy in the FRW metric. Section \ref{The isotropic case} contains our choice for the form of the components of the vector field for them to be in agreement with the symmetries in the FRW metric. We end the section with the dynamical system we will study. In section \ref{Inflationary dynamical analysis} we study the early time dynamics of the extended model $(\ref{Extended})$ proposed in this paper, we find the fixed points of the system using traditional dynamical system analysis and, with the linearized behavior of the equations of motions around the aforementioned fixed points, we study the stability of the system that is later confirmed and extended with numerical calculations. Section \ref{Recovering General Relativity} discusses the dynamical generation of the Planck mass in the extended model $(\ref{Extended})$ finding out that the addition of the vector field does not affect this mass generation. Some light constrains on the couplings of the model are also found in this section. Section \ref{Late time Dynamical analysis} discusses the late time dynamics of a reduced version of the extended model (\ref{Extended}). Finally, section \ref{Conclusions} contains our conclusions.

\section{Addition of a vector field}\label{Addition of a vector field}

The main motivation of this work is the action found in \cite{Rinaldi} but several different scale invariant models have been studied in for example \cite{Higgs-Dilaton Cosmology: From the Early to the Late Universe}-\cite{Quantum scale invariance} 

\begin{equation}\label{Rinaldi}
    S=\int d^{4}x\sqrt{-g}\left[\frac{\alpha}{36} R^{2}+\frac{\beta}{6}\phi^{2}R-\frac{1}{2}(\partial_{\mu}\phi)^{2}-\frac{\lambda}{4}\phi^{4}\right].
\end{equation}

This model can be thought as the minimal scale invariant model composed only of the Ricci scalar $R$ and a single scalar field $\phi$ without operators of $dim>4$, in another words, this model contains all the possible combinations of terms with units $[E^{4}]$ composed of $R$ and $\phi$ with dimensionless coupling constants $\alpha$, $\beta$ and $\gamma$\footnote{No irrelevant couplings are considered.}. Having said this, in order to further generalize the model $(\ref{Rinaldi})$ we will either need to consider irrelevant couplings in order to introduce higher order terms or to consider another type of field keeping only marginal couplings. In this spirit, one can consider the addition of a vector field $A_{\mu}$ in the following form:

\begin{equation}\label{Extended}
    S=\int d^{4}x\sqrt{-g}\left[\frac{\alpha}{36} R^{2}+\frac{\beta}{6}\phi^{2}R-\frac{1}{2}(\partial_{\mu}\phi)^{2}-\frac{\lambda}{4}\phi^{4}-\frac{1}{4}F_{\mu\nu}F^{\mu\nu}+\gamma\phi^{2}A_{\mu}A^{\mu}\right].
\end{equation}
Where the Faraday tensor is given by

\begin{equation}
    F_{\mu\nu}=\nabla_{\mu}A_{\nu}-\nabla_{\nu}A_{\mu},
\end{equation}
the covariant derivative $\nabla_{\mu}$ is defined as $$\nabla_{\mu}A_{\nu}=\partial_{\mu}A_{\nu}-\Gamma_{\ \mu\nu}^{\gamma}A_{\gamma},$$ and $A_{\mu}$ the vector field with components $A_{\mu}=(-\varphi,A_{i})$. Notice that, since we are working with a torsionless connection $\Gamma^{\alpha}_{\ \beta\gamma}=\Gamma^{\alpha}_{\ \gamma\beta}$, the Faraday tensor reduces to 
\begin{equation}
    F_{\mu\nu}=\partial_{\mu}A_{\nu}-\partial_{\nu}A_{\mu}.
\end{equation}

Other versions of vector inflation have already been studied by several authors 
\cite{Vector Inflation}-\cite{Dark energy as a fixed point of the Einstein Yang-Mills Higgs Equations}. A second motivation for the inclusion of a vector field is the fact that, as we will see, this field in principle is incompatible with the cosmological principle and thus, one may explore models as $(\ref{Extended})$ in the so called Bianchi type cosmologies \cite{Anisotropic scalar - tensor cosmologies} or to reduce the freedom of the vector components. These two approaches although very different could lead to some interesting results either by parting ways with the FRW metric or to be forced to consider special conditions on the vector components as we will do in this paper. Before moving on we need to say a few words on the model (\ref{Extended}). In principle we could proceed to analyze the dynamics of the model in the typical FRW background, however as already mentioned, the addition of a vector field, in general, contradicts one of the key assumptions of the FRW metric and so also contradicts the cosmological principle. The fact that the vector field has three spatial components means that, in general, our system is no longer isotropic. 
This conflict can be solved by considering a non isotropic metric of the form 

\begin{equation}
    ds^{2}=-dt^{2}+a_{i}^{2}dx_{i}^{2}.
\end{equation}
Where $a_{i}$ are the three different scale factors in each spatial direction. In this anisotropic metric, the energy momentum tensor for the vector field is given by 

\begin{equation}
      T_{\mu\nu}=F_{\mu}^{\ \rho}F_{\nu\rho}-\frac{1}{4}F_{\rho\sigma}F^{\rho\sigma}g_{\mu\nu}-g_{\mu\nu}V_{int}+2V^{(A)}_{int}A_{\mu}A_{\nu}.
\end{equation}
Where $V^{(A)}_{int}=\frac{\partial}{\partial A^{2}}V_{int}$.
The $00$ and $ii$ components of this energy momentum tensor are 

\begin{equation}
    T_{00}=\frac{1}{2}\sum^{3}_{i=1}\frac{\dot{A}_{i}^{2}}{a_{i}^{2}}+V_{int}+2V^{(A)}_{int}\varphi^{2},
\end{equation}
\begin{equation}
    T_{ij}=-\dot{A}_{i}\dot{A}_{j}+2V^{(A)}_{int}A_{i}A_{j}+a_{i}a_{j}\left[\frac{1}{2}\sum_{k=1}^{3}\frac{\dot{A}_{i}^{2}}{a_{i}^{2}}-V_{int}\right]\delta_{ij}.
\end{equation}
In principle the $00$ and the $ii$ components are the only ones calculated because of the diagonal property of the perfect fluid energy momentum tensor, however, for the vector field we have off-diagonal components such as

\begin{equation}
    T_{0i}=2V^{(A)}_{int}\varphi A_{i}.
\end{equation}
We can immediately notice that, as the anisotropic metric is still diagonal thus are the modified Einstein equations, therefore the off-diagonal components $T_{0i}$ must vanish. Hence, we have three choices, to have a vanishing scalar potential $\varphi=0$, to have vanishing spatial components of the vector field $A_{i}=0$ or to have a vanishing scalar field $\phi=0$.
Having a vanishing scalar field is out of the question because we would then lose many terms of the model (\ref{Extended}), having a vanishing spatial vector field components is a trivial solution, and therefore the best option is to have a vanishing scalar potential $\varphi=0$ hence $A_{\mu}=(0,A_{i})$.
Moreover, the $ij$ components of the energy momentum tensor impose the following relations for the spatial components of the vector field 

\begin{equation}
    2V^{(A)}_{int}A_{i}A_{j}=\dot{A}_{i}\dot{A}_{j}.
\end{equation}
For our specific potential $V_{int}=\gamma\phi^{2}A_{\mu}A^{\mu}$ the energy momentum tensor reads

\begin{equation}
      T_{\mu\nu}=F_{\mu}^{\ \rho}F_{\nu\rho}-\frac{1}{4}F_{\rho\sigma}F^{\rho\sigma}g_{\mu\nu}+\gamma\phi^{2}A_{\mu}A_{\nu}.
\end{equation}
The $00$ and $ii$ components are given by 

\begin{equation}
    T_{00}=\frac{1}{2}\left(\frac{\dot{A}_{x}^{2}}{a_{x}^{2}}+\frac{\dot{A}_{y}^{2}}{a_{y}^{2}}+\frac{\dot{A}_{z}^{2}}{a_{z}^{2}}\right)+\gamma\phi^{2}\frac{A_{i}^{2}}{a_{i}^{2}},
\end{equation}
and 
\begin{equation}
    T_{ii}=-\dot{A}_{i}^{2}+2\gamma\phi^{2}A_{i}^{2}+a_{i}^{2}\left[\frac{1}{2}\left(\frac{\dot{A}_{x}^{2}}{a_{x}^{2}}+\frac{\dot{A}_{y}^{2}}{a_{y}^{2}}+\frac{\dot{A}_{z}^{2}}{a_{z}^{2}}\right)-\gamma\phi^{2}\frac{A_{i}^{2}}{a_{i}^{2}}\right].
\end{equation}
Finally, the condition on the vanishing $ij$ component of the energy momentum tensor reads

\begin{equation}
    2\gamma\phi^{2}=\frac{\dot{A}_{i}\dot{A}_{j}}{A_{i}A_{j}}.
\end{equation}
Another option in order to reconcile vector fields and the supposed isotropy of the universe is to consider that, either all the components but one of the vector field are zero or that they all behave in the same way. 

\section{The isotropic case}\label{The isotropic case}

In this paper we will explore the, so to say, simple solution of reconciliation of vector fields and the isotropy of the FRW metric. We will work in the Lorenz gauge where $\nabla_{\mu} A^{\mu}=0$ noticing that the fact that $A_{0}=0$ was already motivated. In this isotropic case we would even further reduce the components of the vector field having only one non vanishing component 

\begin{equation}\label{Vectorcomponents}
    A_{\mu}=\left(0,0,0,\frac{A_{z}}{a}\right),
\end{equation}
and therefore, with this reduction of the components of the vector field we are now able to  choose the FRW background 

\begin{equation}
    ds^{2}=-dt^{2}+a^{2}(t)dx_{i}dx^{i}.
\end{equation}

Let us briefly discuss the implications on the choice $(\ref{Vectorcomponents})$ we just made. Fist of all, it is obvious that this choice dramatically reduce the degrees of freedom of the vector field, although the temporal part is set to zero for the off diagonal part of the energy momentum tensor to be null, the fact that two of the spatial components are set to zero is a choice we make. Notice that this automatically makes the vector field equations to be not covariant \footnote{We could think of the $z$ component of the vector field as a scalar field in order for it to be covariant.}. However, the most important consequence of the addition of the vector scalar interaction and the choice $(\ref{Vectorcomponents})$ is that the lagrangian is not gauge invariant. This issue has been presented in other articles such as 
\cite{Dark energy as a fixed point of the Einstein Yang-Mills Higgs Equations}-\cite{Cosmological models with Yang-Mills fields}
where the authors explained that, in a sense, the symmetries of the FRW metric \enquote{oversee} or \enquote{override} the gauge symmetry. Another possible solution is simply the fact that the lagrangian (\ref{Extended}) is not gauge invariant is unimportant for our analysis.

The Maxwell  type equation for the model (\ref{Extended}) read 

\begin{equation}
    \nabla_{\mu}F^{\mu\nu}=2\gamma\phi^{2}A^{\nu}.
\end{equation}
It is more convenient to cast these equations in terms of the field as  

\begin{equation}
    \nabla_{\mu}\nabla^{\mu}A^{\nu}-  \nabla_{\mu}\nabla^{\nu}A^{\mu}=2\gamma\phi^{2}A^{\nu}.
\end{equation}
Commuting the second term 

\begin{equation}\label{vecpot}
\begin{split}
    &\Box A^{\nu}-  \nabla^{\nu}\nabla_{\mu}A^{\mu}-[\nabla_{\mu},\nabla^{\nu}]A^{\mu}-2\gamma\phi^{2}A^{\nu}=0,\\
    &\Box A^{\nu}-  \nabla^{\nu}\nabla_{\mu}A^{\mu}-R_{\mu}^{\ \nu}A^{\mu}-2\gamma\phi^{2}A^{\nu}=0,
\end{split}
\end{equation}
and for our specific gauge and form of $A_{\mu}$ we then have 

\begin{equation}\label{VectorEq}
     \frac{\ddot{A}_{z}}{a}+H\frac{\dot{A}_{z}}{a}+\left(2H^{2}+\dot{H}-2\gamma\phi^{2}\right)\frac{A_{z}}{a}=0.
\end{equation}
Not that that this equation is very reminiscent of that of a scalar field in an FRW background. 
Similarly, the Klein Gordon equation for the scalar field is given by 

\begin{equation}\label{KleinGordon}
   \ddot{\phi}+3H\dot{\phi}-2\beta\phi\dot{H}-\phi\left(4\beta H^{2}-\lambda\phi^{2}+2\gamma \frac{A_{z}^{2}}{a^{2}}\right)=0.
\end{equation}
Finally, by taking the variation of $(\ref{Extended})$ with respect to the metric tensor we can, by choosing the $00$ component of the modified Einstein equations, obtain the analogous to the first Friedmann equation which reads

\begin{equation}\label{FriedmanExtended}
    \alpha\left[6H^{2}\dot{H}-\dot{H}^{2}+2H\ddot{H}\right]+\beta\left[H^{2}\phi^{2}+2H\phi\dot{\phi}\right]-\frac{1}{2}\dot{\phi}^{2}-\frac{\lambda\phi^{4}}{4}-\frac{1}{2}\frac{\dot{A}^{2}_{z}}{a^{2}_{z}}-\gamma\phi^{2}\frac{A_{z}^{2}}{a^{2}}=0.
\end{equation}
Where we used the form of the Ricci scalar in a FRW background $R=12H^{2}+6\dot{H}$ where $H=\dot{a}/a$. These three equations $(\ref{VectorEq})$, $(\ref{KleinGordon})$ and $(\ref{FriedmanExtended})$ can be thought as a dynamical system with four variables, $A_{z}$, $a$, $\phi$, and $H$ therefore the system, at first sight, is non autonomous. This problem could in principle be solved by adding a fourth equation but also by defining a new variable $B$ as 

\begin{equation}
    B=\frac{A_{z}}{a}\quad \mbox{hence} \quad  \frac{\dot{A}_{z}}{a}=\dot{B}+HB \quad \mbox{and} \quad \frac{\ddot{A}_{z}}{a}=\ddot{B}+2H\dot{B}+B\left(\dot{H}+H^{2}\right).
\end{equation}
Consequently,  the equations of motion $(\ref{VectorEq})$, $(\ref{KleinGordon})$ and $(\ref{FriedmanExtended})$ in terms of $B$ are given by 

\begin{equation}\label{vector}
     \ddot{B}+3H\dot{B}+\left(4H^{2}+2\dot{H}-2\gamma\phi^{2}\right)B=0,
\end{equation}

\begin{equation}\label{scalar}
   \ddot{\phi}+3H\dot{\phi}-2\beta\phi\dot{H}-\phi\left(4\beta H^{2}-\lambda\phi^{2}+2\gamma B^{2}\right)=0,
\end{equation} 

\begin{equation}\label{ModifiedEinstein}
    \alpha\left[6H^{2}\dot{H}-\dot{H}^{2}+2H\ddot{H}\right]+\beta\left[H^{2}\phi^{2}+2H\phi\dot{\phi}\right]-\frac{1}{2}\dot{\phi}^{2}-\frac{\lambda\phi^{4}}{4}-\frac{1}{2}\dot{B}^{2}-HB\dot{B}-\left(\gamma\phi^{2}+\frac{H^{2}}{2}\right)B^{2}=0,
\end{equation}
which now is an autonomous dynamical system with three variables, $B$, $\phi$ and $H$.

\section{Inflationary dynamical analysis}\label{Inflationary dynamical analysis}

\subsection{Fixed points}

In this subsection we will focus on finding the fixed points of the dynamical system displayed above. For this, and because we are interested in inflationary fixed points, we change variables from cosmic time $t$ to number of $e$-folds $N$. The number of $e$-folds is defined as $N=\ln a$ and thus, the Hubble parameter and its derivatives are given by: 

\begin{equation}\label{Hchain}
    H=\frac{\dot{a}}{a}=\frac{d}{dt}\ln{a}=\frac{dN}{dt},
\end{equation}
\begin{equation*}
    \dot{H}=\frac{dH}{dN}\frac{dN}{dt}=H'H \quad \mbox{and} \quad \ddot{H}=H'^{2}H+H^{2}H''.
\end{equation*}
Similarly, the scalar and vector fields derivatives are rewritten as follows:  

\begin{equation}\label{Pchain}
    \dot{\phi}=\phi'H, \quad \ddot{\phi}=\phi'H'H+H^{2}\phi'', \quad\mbox{and} \quad \dot{B}=B'H, \quad \ddot{B}=B'H'H+H^{2}B''.
\end{equation}
In terms of the $e$-folding number the dynamical system composed of $(\ref{VectorEq})$, $(\ref{KleinGordon})$ and $(\ref{FriedmanExtended})$ takes the form  

\begin{equation}\label{NB}
    B''+B'\left(\frac{H'}{H}+3\right)+B\left(4+2\frac{H'}{H}-2\gamma\frac{\phi^{2}}{H^{2}}\right)=0,
\end{equation}
\begin{equation}\label{NP}
    \phi''+\phi'\left(\frac{H'}{H}+3\right)-2\beta\phi\frac{H'}{H}-\phi\left(4\beta-\lambda\frac{\phi^{2}}{H^{2}}+2\gamma\frac{B^{2}}{H^{2}}\right)=0,
\end{equation}
\begin{equation}\label{NH}
    H''+H'\left(3+\frac{H'}{2H}\right)+\frac{\beta}{2\alpha}\left(\frac{\phi^{2}}{H}+2\phi\frac{\phi'}{H}\right)-\frac{1}{4\alpha}\frac{\phi'^{2}}{H}-\frac{\lambda\phi^{4}}{8\alpha H^{3}}-\frac{1}{4\alpha}\frac{B'^{2}}{H}-\frac{BB'}{2\alpha H}-\left(\frac{\gamma}{2\alpha}\frac{\phi^{2}}{H^{3}}+\frac{1}{4\alpha H}\right)B^{2}=0.
\end{equation}

Now, following standard dynamical analysis techniques, see for example \cite{Dynamical systems}, we take all the derivatives in $(\ref{NB})$, $(\ref{NP})$ and $(\ref{NH})$ equal to zero which leads to the following algebraic equations 

\begin{equation}
    \left(4H^{2}-2\gamma\phi^{2}\right)B=0,
\end{equation}
\begin{equation*}
    \phi\left(4\beta H^{2}-\lambda\phi^{2}+2\gamma B^{2}\right)=0,
\end{equation*}
\begin{equation*}
    \beta\phi^{2}H^{2}-\frac{\lambda}{4}\phi^{4}-\left(\gamma\phi^{2}+\frac{H^{2}}{2}\right)B^{2}=0,
\end{equation*}
whose solutions will then be the fixed points of the dynamical system. It is then easy to see that there are three generic solutions for the above system

\begin{itemize}
\item Fixed point $(a)$\\
    \begin{equation}\label{FPa}(H,\dot{H},\phi,\dot{\phi},B,\dot{B})=\left(\sqrt{\frac{\lambda}{4\beta}}\phi,0,\phi,0,0,0\right).
    \end{equation}
\item Fixed point $(b)$\\
    \begin{equation}\label{FPb}(H,\dot{H},\phi,\dot{\phi},B,\dot{B})=(H,0,0,0,0,0).
    \end{equation}
\item Fixed point $(c)$\\
    \begin{equation}\label{FPc}(H,\dot{H},\phi,\dot{\phi},B,\dot{B})=(0,0,0,0,B,0).
    \end{equation}
\end{itemize}

Before moving on, notice that the fixed point $(\ref{FPa})$ can be found in at least two different ways. One in which the form of $H$ is initially given by $H^{2}=\frac{\gamma}{2}\phi^{2}$ but then one finds the relation between the couplings $\beta\gamma=\frac{\lambda}{2}$ hence recovering the form above $(\ref{FPa})$. The second way to get this fixed point is to solve the system assuming a solution of the type $B=B(\phi)$ however, this path leads to the vector field to be proportional to the following combination of coupling constants $$ B\propto \left(2\beta\gamma-\lambda\right),$$ while the relation $\beta\gamma=\frac{\lambda}{2}$ still holds, therefore ending up with $B=0$ as in $(\ref{FPa})$.

\subsection{Stability}

We now study the stability of these three fixed points. The idea in this section is to investigate, first, whether or not the addition of the two extra terms in $(\ref{Extended})$ allows the model to still be able to replicate cosmic inflation. For this, the system needs an unstable fixed point where the Hubble parameter is almost constant thus granting a quasi exponential expansion and then a stable fixed point where the system will relax giving birth to the Hubble expansion present in the $\Lambda$CDM model. Summarizing, the unstable fixed point is needed for the universe to inflate and to stop inflating and the stable fixed point is needed for the universe to relax and reproduce the Hubble expansion.\footnote{When the system starts relaxing around the stable fixed point, it starts oscillating giving birth to the reheating of the universe. }
In order to study the stability conditions of each of the three above fixed points we will take the linearized solutions around each fixed point and, depending on the behavior of the three variables we will be able to discover whether the system tends to stay or drive away from these fixed point. 

\subsubsection{Unstable fixed points}
Starting with the fixed point $(\ref{FPb})$ we have the linearized equations

\begin{equation}
    B''+3B'+4B=0,
\end{equation}
\begin{equation}
    \phi''+3\phi'-4\beta \phi=0,
\end{equation}
\begin{equation}
    H''+3H'=0,
\end{equation}
with solutions 

\begin{equation}
    B=c_{1}e^{-3N/2}\sin{\frac{\sqrt{7}N}{2}}+c_{2}e^{-3N/2}\cos{\frac{\sqrt{7}N}{2}},
\end{equation}
\begin{equation}
    \phi=c_{3}e^{-1/2\left(\sqrt{16\beta+9}+3\right)N}+c_{4}e^{1/2\left(\sqrt{16\beta+9}-3\right)N},
\end{equation}
\begin{equation}
    H=c_{5}e^{-3N}+c_{6}.
\end{equation}

Both the vector field and the Hubble parameter tend to a constant while the scalar field solution is a combination of increasing and decreasing modes. This behavior is reminiscent of a saddle point and thus an unstable fixed point. \\
Moving on to the fixed point (\ref{FPc}), we have the linearized equations 

\begin{equation}
    B''+3B'+4B=0,
\end{equation}
\begin{equation}
    \phi''+3\phi'-4\beta\phi=0,
\end{equation}
\begin{equation}
    H''+3H'=0.
\end{equation}
These are exactly the same as the ones for the fixed point $(\ref{FPb})$ and thus this is also an unstable fixed point. Note that in both these fixed points, the Hubble parameter tends to a constant and thus the scale factor tends to an exponential growth. 

\subsubsection{Stable fixed points}
We now analyze the fixed point $(\ref{FPa})$, the linearized equations are

\begin{equation}
    B''+3B'=0,
\end{equation}
\begin{equation}
    \phi''+3\phi'=0,
\end{equation}
\begin{equation}
    H''+3H'=0,
\end{equation}
with solutions 

\begin{equation}
    B=c_{1}e^{-3N}+c_{2},
\end{equation}
\begin{equation}
    \phi=c_{3}e^{-3N}+c_{4},
\end{equation}
\begin{equation}
    H=c_{5}e^{-3N}+c_{6}.
\end{equation}
Thus, as the three variables tend to get constant, the system will never drag away from the fixed point and so the fixed point $(\ref{FPa})$ is a stable fixed point. Therefore, dynamically, the model $(\ref{Extended})$ has the appropriate stability to describe cosmic inflation. Moreover, notice that the addition of the vector field only accounted, dynamically, as an additional unstable fixed point that behaves exactly as the one without, or with null, vector field. In this sense, the results in \cite{Rinaldi} for early times (inflation) are unaffected by the addition of the vector field. We suspect this unaffected behavior is a consequence on the choice $(\ref{Vectorcomponents})$ as we could thought of $B$ as a second scalar field minimally coupled to gravity, therefore, for this choice, the model $(\ref{Extended})$ could be thought as a two (scalar) field model with the difference that one scalar field is minimally coupled while the other one is non minimally coupled.\footnote{This is different from traditional multifield inflation models because in such models the two fields tend to have the same equations of motion.}
In fact, the model $(\ref{Extended})$ could be further generalized including not only the vector scalar interaction but also including the following two scale invariant terms 

\begin{equation}\label{additional terms}
    \beta_{1} A_{\mu}A^{\mu}R, \quad \mbox{and} \quad \lambda_{1} A_{\mu}A^{\mu}A_{\nu}A^{\nu}.
\end{equation}
Where $\beta_{1}$ and $\lambda_{1}$ are coupling constants. These two terms should be included in the original action $(\ref{Extended})$ as they are also composed of the vector field $A_{\mu}$ and the Ricci scalar $R$ which are components of the extended model $(\ref{Extended})$. However, notice that in the present analysis, the choice $(\ref{Vectorcomponents})$ effectively reduces the vector field to a new scalar field $B$ and thus the two terms $(\ref{additional terms})$ are accounted by a simple redefinition of the couplings 

\begin{equation}
    \lambda\xrightarrow{}\lambda+\lambda_{1}, \quad \mbox{and} \quad \beta\xrightarrow{}\beta+\beta_{1}.    
\end{equation}
\subsubsection{Numerical Analysis}\label{NumericalAnalysis}

It is clear that the above section is an approximation we made in order to analytically study the dynamics of the  model $(\ref{Extended})$, the linerized equations found above should be a good approximation of the real behavior of each variables but only very near the fixed points of the model. Moreover, from the analytic procedure, it is not clear that the unstable and stable fixed points are indeed connected, the fact that the system possesses these two fixed points does not automatically means that the system can indeed transition from one fixed point to another. The goal of this section is to study the full dynamics of the model $(\ref{Extended})$ for us to verify the analytical results previously found. For this analysis and, as this paper is an extension of \cite{Rinaldi}, we choose the same values as the ones in this reference $\alpha=2250$, $\beta=15$, $\lambda=1/10$ and $\gamma=1/300$ together with the initial conditions $H(0)=1$,  $\phi(0)=\phi'(0)=H'(0)=10^{-8}$ with the only difference being the initial conditions for the vector field $B(0)=B'(0)=100$. 

\begin{figure}[h]
\centering
\begin{subfigure}{.5\textwidth}
  \centering
  \includegraphics[width=1\linewidth]{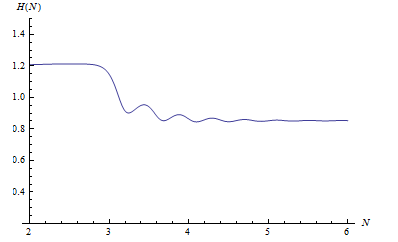}
  \caption{Plot of $H(N)$.}
  \label{He}
\end{subfigure}%
\begin{subfigure}{.5\textwidth}
  \centering
  \includegraphics[width=1\linewidth]{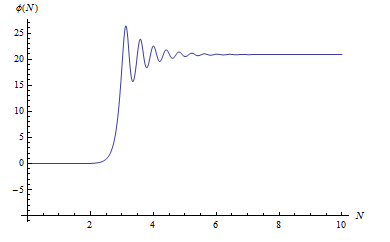}
  \caption{Plot of $\phi (N)$.}
  \label{pe}
\end{subfigure}
\caption{Figure [\ref{He}] and [\ref{pe}] show the behavior of the Hubble parameter $H$ and the scalar field $\phi$ as functions of the number of $e$-folds $N$. Both of these presenting a plateau identified with inflation and an oscillatory phase.}
\label{Hepe}
\end{figure}

\begin{figure}[h]
\centering
\begin{subfigure}{.5\textwidth}
  \centering
  \includegraphics[width=1\linewidth]{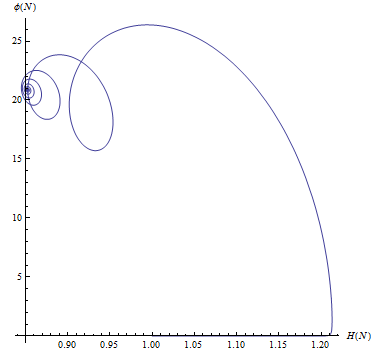}
  \caption{Phase portrait of $\phi(N)$ and $H(N)$.}
  \label{pHe}
\end{subfigure}%
\begin{subfigure}{.5\textwidth}
  \centering
  \includegraphics[width=1\linewidth]{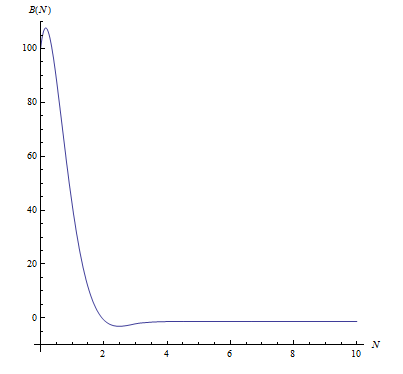}
  \caption{Plot of $B(N)$.}
  \label{Be}
\end{subfigure}
\caption{In figure [\ref{pHe}] we show the phase portrait of the plane $\phi$ $vs$ $H$ where, starting from an unstable fixed point, it is clear the the system is dragged to its stable fixed point. Figure [\ref{Be}] shows the behavior of the vector field $B$ as a function of the number of $e$-folds $N$. Here we see how the vector field is quickly diluted due to the inflationary period.}
\label{pHeBe}
\end{figure}

In FIG $\ref{He}$. we show the behavior of the Hubble parameter noticing that the plateau located at the left side of the figure is identified with the unstable fixed point where the universe inflates. This is the region where the Hubble parameter is constant leading to an exponential expansion. Then, after some oscillations, the Hubble parameters tends to a constant again due to the model $(\ref{Extended})$ not considering any SM content. Note that this plot is valid for just a few oscillations because after that, the $\Lambda$CDM model dynamics comes to play with the non zero value of the Hubble parameter. In FIG \ref{pe} the behavior of the scalar field is presented, this being similar to that of the Hubble parameter, the scalar field also has a plateau where it becomes constant an thus its potential has a flat region where the universe can inflate, then, the scalar field also starts oscillating. 
FIG \ref{pHe}. contains the phase portrait of the scalar field and the Hubble parameter. This figure clearly shows how the system, after some oscillations, reach an stable fixed point. Finally in FIG \ref{Be} we see how the vector field $B$ is washed out very quickly, precisely what we would expect to happen during the inflationary epoch. This was already anticipated in the analytical analysis carried out in the last section as all the local solutions for the vector field tend to a constant, in this case zero. Overall, comparing with the original model in \cite{Rinaldi} we see that the inflationary dynamics are unaffected by the addition of the vector field $B$, at the local (analytical) and numerical level.

\section{Recovering General Relativity}\label{Recovering General Relativity}
As already mentioned, the action (\ref{Extended}) is scale invariant and, as we are interested in early universe solutions, a.k.a inflation, we must ensure that the model (\ref{Extended}) has the capability of recovering Einstein's gravity in order for the universe to possibly undergo the Hot Big Bang expansion. As (\ref{Extended}), at the level of the lagrangian is forever scale invariant, we need to investigate the possibility of dynamically generate a mass scale that could be identified with the Planck mass $M_{p}$. In order to explore this possibility, lets consider the effective potential $V_{eff}$ given by 

\begin{equation}
    V_{eff}=-\frac{\beta}{6}\phi^{2}R+\frac{\lambda}{4}\phi^{4}-\gamma B^{2}\phi^{2}=-\left(\frac{\beta}{6}R+\gamma B^{2}\right)\phi^{2}+\frac{\lambda}{4}\phi^{4}.
\end{equation}
Where the bare mass $m_{0}$ of the field $\phi$ is given by 

\begin{equation}
    m^{2}_{0}=\frac{\beta}{6}R+\gamma B^{2}.
\end{equation}
The first derivative of the effective potential 

\begin{equation}
    \frac{\partial}{\partial\phi}V_{eff}=-2\left(\frac{\beta}{6}R+\gamma B^{2}\right)\phi+\lambda \phi^{3}=\phi\left(-\frac{\beta}{3}R-2\gamma B^{2}+\lambda\phi^{2}\right),
\end{equation}
vanishes at the points 

\begin{equation}
    \phi_{1}=0 \quad \mbox{and} \quad \phi_{2}^{2}=\frac{\frac{R\beta}{3}+2\gamma B^{2}}{\lambda}.
\end{equation}
Notice that the solution $\phi_{2}$ has some limiting cases, namely when $B=0$ and when $R=H=0$. These two scenarios leads to the two fixed points 

\begin{equation}
    \phi_{3}^{2}=\frac{2\gamma B^{2}}{\lambda} \quad \mbox{and} \quad \phi_{4}^{2}=\frac{R\beta}{3\lambda}.
\end{equation}
As now we also have a vector field inside the effective potential, we can also take the derivative 

\begin{equation}
    \frac{\partial}{\partial B}V_{eff}=-2\gamma\phi^{2}B,
\end{equation}
which vanishes at the points $B=0$ and $\phi=0$.
Hence, in principle, the three fixed points $\phi_{2}$, $\phi_{3}$ and $\phi_{4}$ could give rise to symmetry breaking however, we need to recall that the stable fixed point is the one that needs to recover GR. In this spirit then, we need to recall that inflation happens at an unstable fixed point where, traditionally, the scalar field potential is almost constant (flat). Contrary to this, the stable fixed point is the one we need to break the symmetry in, the one that needs to reproduce GR for the system to transition to the radiation dominated epoch. The one stable fixed point we found is $(\ref{FPa})$ thus, the fixed point that will give rise to a dynamically generated mass is given by $\phi_{4}$. At this fixed point, as we need a constant expansion $\dot{H}=0$ then the Ricci scalar is constant and so does is the scalar field, hence $\phi_{4}=\phi_{0}$ with $\phi_{0}$ the constant value of $\phi$ around the fixed point $\phi_{4}$. Consequently, the kinetic term for the scalar field vanishes. Similarly, as this fixed point requires a vanishing vector field $B$, the only remaining terms in the lagrangian are 

\begin{equation}
    \frac{\alpha}{36} R^{2}+\frac{\beta}{6}\phi^{2}R-\frac{\lambda}{4}\phi^{4},
\end{equation}
which, evaluated at the fixed point are 

\begin{equation}
    \frac{\alpha}{36} R^{2}+\frac{\beta}{6}\phi_{0}^{2}R-\frac{\beta^{2}R^{2}}{36\lambda}=\frac{\beta}{6}\phi_{0}^{2}R,
\end{equation}
with $\alpha=\beta^{2}/\lambda$\footnote{Note that this relation is exactly the same found in the Einstein frame.}. Finally, it is natural to identify the remaining term with the Einstein Hilbert term

\begin{equation}
    \frac{1}{2}M_{p}^{2}R=\frac{\beta}{6}\phi_{0}^{2}R,
\end{equation}
so, the dynamically generated Planck mass is set to be 

\begin{equation}
    M_{p}^{2}=\frac{\beta}{3}\phi_{0}^{2}.
\end{equation}

\begin{figure}[h]
    \centering
    \includegraphics[width=.6\linewidth]{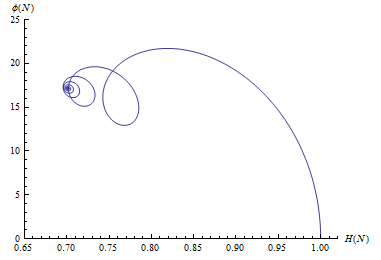}
    \caption{Phase portrait of $\phi(N)$ and $H(N)$ with identical numerical values as those in \cite{Rinaldi}. This plot confirms how the addition of the vector field $B$ does not affect the stable fixed point of the system and therefore does not affect the dynamical generation of the Planck mass.}
    \label{pHGR}
\end{figure}

Note also that, even with the addition of the vector field, the generic dynamic of the system is still the same as in the original paper from Rinaldi and collaborators \cite{Rinaldi}, the early universe solutions are still present and unaffected by the vector field. This is very clear when considering the derivative of the effective potential with respect to $B$, as the only  vanishing points are located at zero, there is no room for symmetry breaking and thus no way to recover GR using the vector field. This, in principle does not mean the vector field is unable to be the responsible for inflation but that the dynamical generation of the Planck mass it still determined by the non zero vaccum of the scalar field. The fact that the dynamical generation of the Planck mass is unaffected by the vector field can be further verified by taking the exact same numerical values  and initial conditions as those in \cite{Rinaldi} and taking $B(0)=B'(0)=10^{-3}$ similarly to those of the scalar field $\phi$. The phase portrait of $\phi(N)$ and $H(N)$ shown in FIG \ref{pHGR}. is unaffected by the addition of the vector field $B$. This means that the stable fixed point, the one we break the symmetry in,
is exactly the same one as in \cite{Rinaldi}.

\subsubsection{Constrains on the couplings}

As already explained, inflation occurs at an unstable fixed point, a point the systems tends to go to but one that the system is doomed to apart from. In our present system we have, in principle, two unstable fixed points $(H,\dot{H},\phi,\dot{\phi},B,\dot{B})=(H,0,0,0,0,0)$ and $(H,\dot{H},\phi,\dot{\phi},B,\dot{B})=(0,0,0,0,B,0)$ however, notice that the local  behavior of $H$, $\phi$ and $B$ are the same for both fixed points. Around these unstable fixed points we can expand our variables as 

\begin{equation}\label{Expansions}
    H=H_{0}, \quad \phi=\phi_{0}e^{\Delta N}, \quad \mbox{and} \quad B=B_{0}.
\end{equation}
Where $\Delta N=N_{f}-N_{i}$ is the number of $e$-folds up to the end of inflation. Therefore, starting from 

\begin{equation}
    \phi''+\phi'\left(\frac{H'}{H}+3\right)-2\beta\phi\frac{H'}{H}-\phi\left(4\beta-\lambda\frac{\phi^{2}}{H^{2}}+2\gamma\frac{B^{2}}{H^{2}}\right)=0,
\end{equation}
we can impose the slow roll condition on the Hubble factor $$-\frac{\dot{H}}{H^{2}}=-\frac{H'}{H}<<1.$$
Thus having

\begin{equation}
    -\frac{H'}{H}=\frac{\phi''-\phi\left(4\beta-\lambda\frac{\phi^{2}}{H^{2}}+2\gamma\frac{B^{2}}{H^{2}}\right)+3\phi'}{\phi'-2\beta\phi}.
\end{equation}
Setting the end of inflation around the time when $$-\frac{H'}{H}\approx 1,$$ and expanding the variables as in $(\ref{Expansions})$ we have 

\begin{equation}
    1=\frac{\phi_{0}e^{\Delta N}-\phi_{0}e^{\Delta N}\left(4\beta-\lambda\frac{e^{2\Delta N}\phi_{0}^{2}}{H_{0}^{2}}+2\gamma\frac{B_{0}^{2}}{H_{0}^{2}}\right)+3e^{\Delta N}\phi_{0}}{e^{\Delta N}\phi_{0}-2\beta e^{\Delta N}\phi_{0}}.
\end{equation}
Which solution for $\Delta N$ is given by 

\begin{equation}\label{Numberefolds}
    \Delta N=\frac{1}{2}\ln\left[\frac{H_{0}^{2}}{\lambda \phi_{0}^{2}}\left(-3+2\beta\right)+\frac{\gamma}{\lambda}\frac{B_{0}^{2}}{\phi_{0}^{2}}\right].
\end{equation}
Note that $(\ref{Numberefolds})$ reduces to the one found in \cite{Rinaldi} in the limit of vanishing coupling $\gamma$. Furthermore, even for $\beta=\gamma=0$ $(\ref{Numberefolds})$ still survives because the model $(\ref{Extended})$, in this limit, is just a version of the Starobisnky model. We can also immediately set some constrains on the couplings of the model by imposing that the argument of the logarithm is positive

\begin{equation}\label{coupling condition}
    \beta>-\frac{\gamma B_{0}^{2}}{2H_{0}^{2}}-\frac{3}{2},
\end{equation}
which is compatible with the requirement of $\beta>0$ in the dynamical generation of the Planck mass.

\section{Late time Dynamical analysis}\label{Late time Dynamical analysis}

At late times the inflaton field $\phi$ is supposed to be at its minimum and hence its energy density is thought to be negligible hence, one could think of a late time scenario of the model $(\ref{Extended})$ to be 

\begin{equation}\label{ReducedExtended}
    S=\int d^{4}x\sqrt{-g}\left[\frac{\alpha}{36} R^{2}-\frac{1}{2}(\partial_{\mu}\phi)^{2}-\frac{1}{4}F_{\mu\nu}F^{\mu\nu}+\gamma\phi^{2}A_{\mu}A^{\mu}\right].
\end{equation}
With equations of motion 

\begin{equation}
    B''+B'\left(\frac{H'}{H}+3\right)+B\left(4+2\frac{H'}{H}-2\gamma\frac{\phi^{2}}{H^{2}}\right)=0,
\end{equation}
\begin{equation}
    \phi''+\phi'\left(\frac{H'}{H}+3\right)-2\phi\gamma\frac{B^{2}}{H^{2}}=0,
\end{equation}
\begin{equation}
\begin{split}
    H''+H'\left(3+\frac{H'}{2H}\right)-\frac{1}{4\alpha}\frac{\phi'^{2}}{H}-\frac{1}{4\alpha}\frac{B'^{2}}{H}-\frac{BB'}{2\alpha H}-\left(\frac{\gamma}{2\alpha}\frac{\phi^{2}}{H^{3}}+\frac{1}{4\alpha H}\right)B^{2}=0.
\end{split}
\end{equation}
Once again, in order to find the fixed points of the new system, we set all derivatives to zero obtaining the algebraic equations  

\begin{equation}\label{1}
    \left(4-2\gamma\frac{\phi^{2}}{H^{2}}\right)B=0,
\end{equation}
\begin{equation}\label{2}
    2\phi\gamma \frac{B^{2}}{H^{2}}=0,
\end{equation}
\begin{equation}\label{3}
    \left(\frac{\gamma}{2\alpha}\frac{\phi^{2}}{H^{3}}+\frac{1}{4\alpha H}\right)B^{2}=0.
\end{equation}
and thus, the fixed points of the system are given by 

\begin{itemize}
    \item Fixed point $(d)$\\
\begin{equation}\label{FPd}
    (H,\dot{H},\phi,\dot{\phi},B,\dot{B})=(H,0,\phi,0,0,0).
\end{equation}
    \item Fixed point $(e)$\\
\begin{equation}\label{FPe}
    (H,\dot{H},\phi,\dot{\phi},B,\dot{B})=(0,0,0,0,B,0).
\end{equation}
\item Fixed point $(f)$\\
\begin{equation}\label{FPf}
    (H,\dot{H},\phi,\dot{\phi},B,\dot{B})=(0,0,\phi,0,0,0).
\end{equation}
\item Fixed point $(g)$\\
\begin{equation}\label{FPg}
    (H,\dot{H},\phi,\dot{\phi},B,\dot{B})=\left(\sqrt{\frac{\gamma}{2}}\phi,0,\phi,0,0,0\right).
\end{equation}
\end{itemize}
The local behavior around the fixed point $(\ref{FPg})$ is given by 

\begin{equation}
    B''+3B'=0 \xrightarrow{} B=c_{1}e^{-3N}+c_{2},
\end{equation}
\begin{equation}
    \phi''+3\phi'=0 \xrightarrow{} \phi=c_{3}e^{-3N}+c_{4},
\end{equation}
\begin{equation}
    H''+3H'=0 \xrightarrow{} H=c_{5}e^{-3N}+c_{6}.
\end{equation}
Similarly, it is easy to show that the local behaviors around the fixed points $(\ref{FPd})$, $(\ref{FPe})$ and (\ref{FPf}) are the same ones and are given by 

\begin{equation}
    B''+3B'+4B=0,
\end{equation}
\begin{equation}
    \phi''+3\phi'=0,
\end{equation}
\begin{equation}
    H''+3H'=0,
\end{equation}
with solutions 
\begin{equation}
    B=c_{1}e^{-3N/2}\sin{\frac{\sqrt{7}N}{2}}+c_{2}e^{-3N/2}\cos{\frac{\sqrt{7}N}{2}},
\end{equation}
\begin{equation}
    \phi=c_{3}e^{-3N}+c_{4},
\end{equation}
\begin{equation}
    H=c_{5}e^{-3N}+c_{6}.
\end{equation}
All the fixed points in this new system are stable. This is congruent with our previous results because in the full model (\ref{Extended}) the coupling $\beta$ made the scalar field to have increasing and decreasing modes. Note also that in every fixed point, the solution for the Hubble parameter is 

\begin{equation}
    H=c_{1}+c_{2}e^{-3N}.
\end{equation}
By recalling that, $N=\ln a$, we have the solution for the scale factor in terms of cosmic time 

\begin{equation}
    a(t)=\left(\frac{e^{3c_{1}t}}{c_{2}}-\frac{c_{2}}{c_{1}}\right)^{1/3},
\end{equation}
which, for late times become 

\begin{equation}
    a(t)=a_{0}e^{c_{1}t},
\end{equation}
and thus a perfect de Sitter fixed point universe as the one we live in today. Following this logic, we can find the late time solution for the scalar and vector fields. The scalar field solution, in terms of cosmic time is 

\begin{equation}
    \phi(t)=c_{3}+c_{4}\left(\frac{1}{e^{3c_{1}t}-\frac{c_{2}}{c_{1}}}\right). 
\end{equation}
Which, for late times becomes constant and so, the scalar field can be identified with the cosmological constant that dominates the universe at late times. Finally, the vector field has two possibilities, either 

\begin{equation}
    B(t)=c_{5}+c_{6}\left(\frac{1}{e^{3c_{1}t}-\frac{c_{2}}{c_{1}}}\right), 
\end{equation}
or

\begin{equation}
    B(t)\propto c_{7}\left(\frac{1}{\frac{e^{3c_{1}t}}{c_{2}}-\frac{c_{2}}{c_{1}}}\right).
\end{equation}
Therefore at late times, in the former case, the vector field also becomes constant and so could be summed up with the scalar field to generate the cosmological constant and in the latter, the vector field is diluted just as in the inflationary epoch. Notice that, the addition of the vector field $B$ is of crucial importance for the previous results. If this field wasn't there, the fixed points equations $(\ref{1})$, $(\ref{2})$ and $(\ref{3})$ would all vanish. 

\subsubsection{Numerical Analysis}

\begin{figure}[h]
\centering
\begin{subfigure}{.5\textwidth}
  \centering
  \includegraphics[width=1\linewidth]{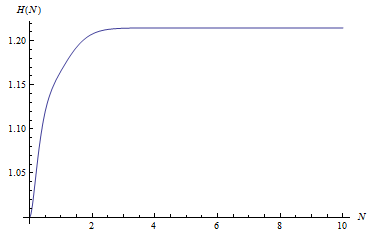}
  \caption{Plot of $H(N)$.}
  \label{Hl}
\end{subfigure}%
\begin{subfigure}{.5\textwidth}
  \centering
  \includegraphics[width=1\linewidth]{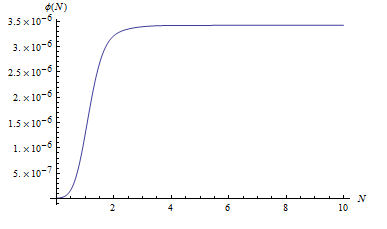}
  \caption{Plot of $\phi (N)$.}
  \label{pl}
\end{subfigure}
\caption{Figure [\ref{Hl}] and [\ref{pl}] show the behavior of the Hubble parameter $H$ and the scalar field $\phi$ as functions of the number of $e$-folds $N$. Both of this presenting a plateau at late times (big $N$) which corresponds to a de Sitter fixed point.}
\label{Hepe}
\end{figure}

We repeat the numerical analysis for the late time model $(\ref{ReducedExtended})$  with the same values and initial conditions as those in $(\ref{NumericalAnalysis})$ finding that both the Hubble parameter FIG \ref{Hl}. and the scalar field FIG \ref{pl}. become constant for late time (big $N$) leading to an exponential solution for the scale factor and a constant value for the scalar field identified with the cosmological constant. This is the exact behavior we found employing analytical methods. 

\begin{figure}[h]
    \centering
    \includegraphics[width=.6\linewidth]{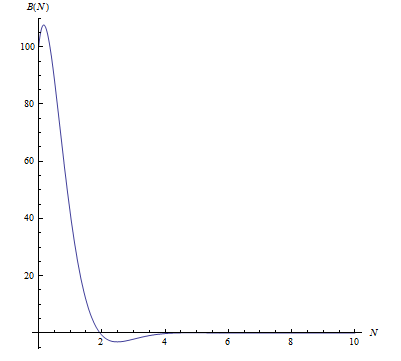}
    \caption{Plot of $B(N)$.}
    \label{Bl}
\end{figure}

In FIG \ref{Bl} we show the behavior of the vector field which is once again diluted very quickly due to the universe being in a de Sitter phase and thus, in principle, for our choice of parameters we have that $c_{5}=0$.

%%%%%%%%%%%%%%%%%%%%%%%%%%%%%%%%%%%%%%%%%%%%%%%%%%%%
%%%%%%%%%%%%%%%%%%%%%%%%%%%%%%%%%%%%%%%%%%%%%%%%%%%%
\section{Conclusions}\label{Conclusions}

We study the model originally proposed in \cite{Rinaldi} with the inclusion of a vector field with a vector scalar interaction. This addition was motivated by the fact that the original model already contained all the possible scale invariant terms containing only the Ricci curvature and a single scalar field with dimensionless couplings. After this addition was motivated, we discuss the generic incompatibility between the vector field and the cosmological principle. We study the isotropic case of the vector field in which only the $z$ component is non vanishing. Following such discussion we find the equations of motion for the extended model and proceed to study its dynamics in an approximate analytical way using traditional dynamical systems methods. We find out that there exist two fixed points corresponding to those of the original paper \cite{Rinaldi} and one extra unstable fixed point born from the addition of the vector field.
After the stability analysis is carried out we conclude that the extended model has the same early times dynamic as the one in \cite{Rinaldi} and thus the vector field does not affect the early time stability of the system, it just dilutes as expected during the inflationary period. We confirm this findings by performing a numerical analysis on the extended model showing that the full dynamics of the original model in \cite{Rinaldi} are unaffected at the local (analytical) and numerical level. We proceed by investigating how the addition of the vector field may affect the dynamical generation of the Planck mass finding out,  and confirmed with a numerical calculation, that this is also unaffected by the presence of the vector field. As a conclusion of the early times, we calculate the number of $e$-folds up to the end of inflation imposing some light constrains on the couplings of the extended model. 
Arguing that for late times some of the terms in the extended model vanishes, we study a reduced version of the model proposed in this paper finding out that this reduced version contains only stable fixed points that, for late times, are exactly de Sitter fixed points with the scalar field becoming constant and identified with the cosmological constant while the vector field is diluted once more. We solve the reduce system using numerical tools and verify this results. 

It is clear that the results in this paper are based on the choice we made on the components of the vector field. Having an isotropic vector field as the one in this paper just effectively contributes as a second minimally coupled scalar field. A natural continuation of this work may be to consider a non isotropic metric and explore the full vector field together with the scale invariant model. 

\acknowledgments
I would like to thank the National Council of Science and Technology (CONACyT) for its funding and support. Moreover I want to thank my advisor Professor G. German for reading and commenting this manuscript and for his support and encouragement. Finally I also want to thank Professor M. Rinaldi for his guidance during my master's studies where I got to know the work that motivated this paper.

%%%%%%%%%%%%%%%%%%%%%%%%%%%%%%%%


\begin{thebibliography}{10}

\bibitem{The Foundation of the General Theory of Relativity}
Einstein, Albert, \emph{The Foundation of the General Theory of Relativity}, \emph{Annalen Phys.} {\bf vol 49} (1916) pg 769-822.


\bibitem{Dynamics of dark energy}
Copeland, Edmund J. and Sami, M. and Tsujikawa, Shinji, \emph{Dynamics of dark energy}, \emph{Int. J. Mod. Phys. D} {\bf vol 15} pg 1753-1936 (2006).

\bibitem{Quintessence, cosmic coincidence, and the cosmological constant}
Zlatev, Ivaylo and Wang, Li-Min and Steinhardt, Paul J. \emph{Quintessence, cosmic coincidence, and the cosmological constant}, \emph{Phys. Rev. Lett.} {\bf vol 82} pg 896-899 (1999).

\bibitem{Coupled quintessence}
Amendola, Luca, \emph{Coupled quintessence}, \emph{Phys. Rev. D} {\bf vol 62} pg 043511 (2000).

\bibitem{An Alternative to quintessence}
Kamenshchik, Alexander Yu. and Moschella, Ugo and Pasquier, Vincent, \emph{An Alternative to quintessence}, \emph{Phys. Lett. B} {\bf vol 511} pg 265-268 (2001).

\bibitem{Cosmological Consequences of a Rolling Homogeneous Scalar Field}
Ratra, Bharat and Peebles, P. J. E. \emph{Cosmological Consequences of a Rolling Homogeneous Scalar Field}, \emph{Phys. Rev. D} {\bf vol 37} pg 3406 (1998).

\bibitem{kinflation}
Armendariz-Picon, C. and Damour, T. and Mukhanov, Viatcheslav F. \emph{k - inflation}, \emph{Phys. Lett. B} {\bf vol 458} pg 209-218 (1999).

\bibitem{Chaotic Inflation}
Linde, Andrei D. \emph{Chaotic Inflation}, \emph{Phys. Lett. B} {\bf vol 129} pg 177-181 (1983).

\bibitem{Hybrid inflation}
Linde, Andrei D. \emph{Hybrid inflation}, \emph{Phys. Rev. D} {\bf vol 49} pg 748-754 (1994).

\bibitem{Guth}
Guth, Alan H.\emph{The Inflationary Universe: A Possible Solution to the Horizon and Flatness Problems}, \emph{Phys. Rev. D} {\bf vol 23} pg 347-356 (1981).

\bibitem{Planck 2018 results. X. Constraints on inflation}
Akrami, Y. and others, \emph{Planck 2018 results. X. Constraints on inflation}, \emph{Astron. Astrophys.} {\bf vol 641} pg A10 (2020).

\bibitem{Starobinsky} A. A. Starobinsky, in Quantum Gravity, Proceedings of the
    2nd Seminar on Quantum Gravity, Moscow, 1981(INR
    Press, Moscow, 1982), pp. 58–72; reprinted inM. A.
    Markov and P. C. West eds., Quantum Gravity(Plenum
    Press, New York, 1984), pp. 103–128; A. A. Starobinsky,
    Phys. Lett.91B, 99102 (1980).

\bibitem{f(R)}
De Felice, Antonio and Tsujikawa, Shinji. \emph{f(R) theories}, \emph{Living Rev. Rel.} {\bf vol 13} pg 3 (2010).

\bibitem{Racioppi}
Gialamas, Ioannis D. and Karam, Alexandros and Pappas, Thomas D. and Racioppi, Antonio and Spanos, Vassilis C. \emph{Scale-invariance, dynamically induced Planck scale and inflation in the Palatini formulation}, \emph{J. Phys. Conf. Ser.} {\bf vol 2105} pg 012005 (2021).

\bibitem{Inflation and reheating in scaleinvariant scalartensor gravity}
Tambalo, Giovanni and Rinaldi, Massimiliano, \emph{Inflation and reheating in scale-invariant scalar-tensor gravity}, \emph{Gen. Rel. Grav.} {\bf vol 49} pg 52 (2017).

\bibitem{Rinaldi}
Rinaldi, Massimiliano and Vanzo, Luciano, \emph{Inflation and reheating in theories with spontaneous scale invariance symmetry breaking}, \emph{Phys. Rev. D} {\bf vol 94} pg 024009 (2016).

\bibitem{Scale invariance}
Gialamas, Ioannis D. and Karam, Alexandros and Pappas, Thomas D. and Racioppi, Antonio and Spanos, Vassilis C. \emph{Scale-invariance, dynamically induced Planck scale and inflation in the Palatini formulation}, \emph{J. Phys. Conf. Ser.} {\bf vol 2105} pg 012005 (2021).

\bibitem{Higgs-Dilaton Cosmology: From the Early to the Late Universe}
Garcia-Bellido, Juan and Rubio, Javier and Shaposhnikov, Mikhail and Zenhausern, Daniel. \emph{Higgs-Dilaton Cosmology: From the Early to the Late Universe}, \emph{Phys. Rev. D.} {\bf vol 84} pg 123504 (2011).

\bibitem{Scale-independent inflation}
Ferreira, Pedro G. and Hill, Christopher T. and Noller, Johannes and Ross, Graham G. \emph{Scale-independent $R^2$ inflation}, \emph{Phys. Rev. D} {\bf vol 100} pg 123516 (2019).

\bibitem{A comparison between the Jordan and Einstein Frames in Brans-Dicke theories with torsion}
Gonzalez Quaglia, R. and German, Gabriel. \emph{A comparison between the Jordan and Einstein Frames in Brans-Dicke theories with torsion}, arXiv. 2206.14228 (2019).

\bibitem{Higgs-Dilaton inflation in Einstein-Cartan gravity}
Piani, Matteo and Rubio, Javier. \emph{Higgs-Dilaton inflation in Einstein-Cartan gravity}, \emph{JCAP} {\bf vol 05} pg 009 (2022).


\bibitem{Quantum scale invariance}
Shaposhnikov, Mikhail and Zenhausern, Daniel. \emph{Quantum scale invariance, cosmological constant and hierarchy problem}, \emph{Phys. Lett. B} {\bf vol 671} pg 162-166 (2009).

\bibitem{Vector Inflation}
Golovnev, Alexey and Mukhanov, Viatcheslav and Vanchurin, Vitaly. \emph{Vector Inflation}, \emph{JCAP} {\bf vol 06} pg 009 (2008).

\bibitem{Vector Field Models of Inflation and Dark Energy}
Koivisto, Tomi and Mota, David F. \emph{Vector Field Models of Inflation and Dark Energy}, \emph{JCAP} {\bf vol 08} pg 021 (2008).

\bibitem{INFLATION DRIVEN BY A VECTOR FIELD}
Ford, L. H. \emph{Inflation driven by a vector field}, \emph{Phys. Rev. D} {\bf vol 40} pg 967 (1989).

\bibitem{Inflation with a massive vector field nonminimally coupled to gravity}
Bertolami, O. and Bessa, V. and P\'aramos, J. \emph{Inflation with a massive vector field nonminimally coupled to gravity}, \emph{Phys. Rev. D} {\bf vol 93} pg 064002 (2016).

\bibitem{Inflation driven by massive vector fields with derivative self-interactions}
Oliveros, A. and Jaraba, Marcos A. \emph{Inflation driven by massive vector fields with derivative self-interactions}, \emph{Int. J. Mod. Phys. D} {\bf vol 28} pg 1950064 (2019).

\bibitem{Cosmological electromagnetic fields and dark energy}
Beltran Jimenez, Jose and Maroto, Antonio L. \emph{Cosmological electromagnetic fields and dark energy}, \emph{JCAP} {\bf vol 03} pg 016 (2009).

\bibitem{Inflation and late-time cosmic acceleration in non-minimal Maxwell-$F(R)$ gravity and the generation of large-scale magnetic fields}
Bamba, Kazuharu and Odintsov, Sergei D. \emph{Inflation and late-time cosmic acceleration in non-minimal Maxwell-$F(R)$ gravity and the generation of large-scale magnetic fields}, \emph{JCAP} {\bf vol 04} pg 024 (2008).

\bibitem{Inflationary cosmology and the late-time accelerated expansion of the universe in non-minimal Yang-Mills-F(R) gravity and non-minimal vector-F(R) gravity}
Bamba, Kazuharu and Nojiri, Shin'ichi and Odintsov, Sergei D. \emph{Inflationary cosmology and the late-time accelerated expansion of the universe in non-minimal Yang-Mills-F(R) gravity and non-minimal vector-F(R) gravity}, \emph{Phys. Rev. D} {\bf vol 77} pg 123532 (2008).

\bibitem{Dark energy as a fixed point of the Einstein Yang-Mills Higgs Equations}
Rinaldi, Massimiliano. \emph{Dark energy as a fixed point of the Einstein Yang-Mills Higgs Equations}, \emph{JCAP} {\bf vol 10} pg 023 (2015).

\bibitem{Anisotropic scalar - tensor cosmologies}
Mimoso, Jose P. and Wands, David. \emph{Anisotropic scalar - tensor cosmologies}, \emph{Phys. Rev. D} {\bf vol 52} pg 5612-5627 (1995).

\bibitem{Isotropy theorem for cosmological Yang-Mills theories}
Cembranos, J. A. R. and Maroto, A. L. and N\'u\~nez Jare\~no, S. J. \emph{Isotropy theorem for cosmological Yang-Mills theories}, \emph{Phys. Rev. D} {\bf vol 87} pg 043523 (2013).

\bibitem{Cosmological models with Yang-Mills fields}
Gal'tsov, Dmitry V. and Davydov, Evgeny A. \emph{Cosmological models with Yang-Mills fields}, \emph{Proc. Steklov Inst. Math.} {\bf vol 272} pg 119-140 (2011).

\bibitem{Dynamical systems}
Bahamonde, Sebastian and B\"ohmer, Christian G. and Carloni, Sante and Copeland, Edmund J. and Fang, Wei and Tamanini, Nicola. \emph{Dynamical systems applied to cosmology: dark energy and modified gravity}, \emph{Phys. Rept.} {\bf vol 775-777} pg 1-122 (2018).
 \end{thebibliography}
\end{document}